\documentstyle[aps]{revtex}

\begin{document}

\title{
        What Are Stories Made Of?
- Quantitative Categorical Deconstruction of Creation
}
\author{
Y. Stolov$^{1}$, M. Idel$^{1}$ 
 and S. Solomon$^{2}$
}
\address{
$^{1}$
 Institute of Jewish Studies, The Hebrew University, Jerusalem
91905, Israel
}
\address{
$^{2}$
Racah Institute of Physics, The Hebrew University, Jerusalem
91904, Israel
}

\maketitle

\begin{abstract}
\newline{}
We extend the Microscopic Representation approach to the quantitative
study of religious and folk stories: A story encrypting symbolically the
creation is deconstructed into its simplest conceptual elements and their
relationships. We single out a particular kind of relationship which we
call "diagonal (or transitive) link": given 2 relations between the
couples of elements AB and respectively BC, the "diagonal link" is the
(composite) relation AC. We find that the diagonal links are strongly and
systematically correlated with the events in the story that are considered
crucial by the experts. We further compare the number of diagonal links in
the symbolic creation story with a folk tale, which ostensibly narrates
the same overt succession of events (but without pretensions of encrypting
additional meanings). We find that the density of diagonal links per word
in the folk story is lower by a factor of 2. We speculate that, as in
other fields the simple transitive operations acting on elementary objects
are at the core of the emergence and recognition of macroscopic meaning
and novelty in complex systems.\\
\end{abstract}


\section{Microscopic Representation and Universal Dynamics}
Several years ago, the Microscopic Representation Approach 
\cite{ref1}
 was introduced in order to understand in a systematic manner the way in which
the human concept describes a very big number of microscopic objects, in
terms of a limited number of macroscopic characters defined by them.
According to this approach, the microscopic picture and the macroscopic
one are considered to be two different pictures of one reality, that are
complimentary to each other, as the ocean is both an aggregate of
molecules and a mass of waves.  In each scale should be found the
macroscopic objects and the laws of their behavior, which are generated
from the microscopic scale.
        At the same period of time a similar approach, the Universal
Dynamics Approach 
\cite{ref2,ref3,ref4} 
was developed but from the opposite direction: from
the universal to the macroscopic. It put the emphasis on the theory,
therefore it is more comfortable for the presentations of the
fundamentals. This approach looked for the most fundamental physical
theory, under the rule that a theory is the more fundamental the less
structure it assumes a priori. It starts with the basic assumption that:
{\bf The human mind thinks about relations between Objects} (this assumption is
becoming more precise after introducing axiomatic characters for
"relations" and "objects"). A collection of objects with their
inter-relations is defined as "a system". Objects of a system can be
systems themselves, i.e. with internal structure, which allows a framework
for the discussion of self-organization in complex systems. The relations
are directed and will therefore be called "arrows", some of which are
identified as fundamental in a given level of description and will be
called "links". Arrows are composed of links and own the main importance
in the description of a system. For example: in a system which is a thin
brick wall - the objects are the bricks and the arrows define the
relations between each brick and its six nearest neighbors, twelve next
nearest neighbors etc. Fig 1 describes a system which is a thin brick wall.
The arrows are those who are defining the structure and turn the wall into
more than a pile of bricks.
        In this framework the dynamics of the system is described: the
basic options of getting a new system from a given one are - death of an
object or a link, duplication of an object or a link, merging identical
objects or links, creating missing adjacent links or promoting an arrow,
which can be composed of two or several links, to the status of link. The
dynamical processes assembled from such elementary operations were called
"drama".

\section{Applications}
        Several applications 
\cite{ref5,ref6,ref7} of the approach to domains outside
natural sciences were developed. Two examples for these applications are
the works in New Product Development \cite{ref8,ref9} and in Creative Advertising Ideas \cite{ref10,ref11}.
In both these domains we do not expect to find well-defined lines of
thought. Yet, during their study according to the universal approach it
was revealed that out of the many possible routes of the dynamics,
specific routes can be clearly defined and shown to be preferred - these
are the "Creativity Templates". The templates can be identified and
described in an abstract and simple way so they can be easily and
efficiently applied and objectively tested. An example for such a template
is the development of a delivery motorbike that bakes pizza during the
delivery ride according to the Component Connection Template (table 1 shows an example for the Component
Connection Template).
Following the simple instructions of the algorithm, leads the dynamics of
the mind to the new creation. 
        The efficiency of the templates was tested quantitatively: a group
of marketing
personal were trained for 4-5 hours in one of the templates. This group
and a control group who did not receive the training were asked to
generate ideas for new products in three categories. The first group used
the template and the second group used conventional ideation methods
(brainstorming, questionnaires etc.). Eight senior marketing professionals
acted as judges and rated the ideas on a 1-7 scale according to their
originality (1=not original at all, 7=very original) and according to
their recommendation to implement the idea (1=not recommended at all,
7=highly recommended). The judges were blind to the identity of the
individuals who generated the ideas and to one another. In spite of the
short length of the training the advantage of the first group was salient:
they received the average of 5.09 vs. 4.18 in originality and 4.40 vs.
3.36 in value.

\section{The Diagonal Link or: The Transitive Template}
        In the present paper we propose a more basic elementary building
block underlying most (may be all) of the creative acts. We call it the
"Diagonal Link" according to its structure or "Transitive Template"
according to its function.
The "Diagonal Link" seems to be the basic element that humans use in
constructing and understanding reality in the widest sense.
In order to test this possibility we use the "Diagonal Link" as the
elementary microscopic operation in the Microscopic Representation study
of Creation as described in symbolic religious stories.
Structurally, the "Diagonal Link" resembles a transitivity connection (fig 2 illustrates transitivity connection).
In the example brought in the previous section it can be seen clearly in
fig 3.
In the same way we found that each of the Creativity Templates included,
at least once, a Diagonal Link as the core of the creative process.
        As the basic assumption of the Microscopic Representation and
Universal Dynamics related to the fundamental infrastructure of the human
way of thinking, we assumed that we will be able to find similar phenomena
in other fields and turned to study the tales of Rabbi Nachman of Bratslav
using this approach. We focused on the first story and studied the
appearances of the Diagonal Link and their role. It should be emphasized
that the analysis was performed in a mechanical method that referred only
to the syntactic structure and not to the meaning of the text.

\section{Rabbi Nachman of Bratslav's Stories and the Creation Story}
        In a very simplified and short way we can say that according to
the Kabala (the Jewish medieval mystical tradition) creation starts when
G-d withdrew in order to make room for the world to be created. After the
creation, a line of light was sent from the infinite G-d into the created
world but due to the deficiencies of the world it could not bear the light
and was "broken" with the sparks of light scattered all around it. The
human task is to collect these divine sparks and bring them back to their
roots. This task is laid upon every human being, more on the Nation of
Israel and most on the righteous person - the Tzadik. This dynamics has
many manifestations: the exile of the soul in the body and the need of the
individual to redeem it by bettering oneself and lifting oneself
spiritually, the exile of the Nation of Israel and their need to return to
the Land of Israel, the exile of the Divine Presence and the need to help
it to be re-united with G-d etc. 
        Rabbi Nachman of Bratslav (1772-1810) saw himself as a Tzadik and
as such he felt, like other Hassidic (the Hassidic movement is a
pre-modern Jewish revival movement, deeply inspired by Kabala) sages,
responsible for the redemption process and for the recruitment of his
followers to help in the process. At a certain point he felt that the
direct cognitive approach did not give enough fruits and that his explicit
teachings did not enlighten his people as he expected. At that point he
decided to convey his message in an implicit manner through story telling.
        The first story that he told was "The Story of The Lost Princess"
\cite{ref13,ref14}. It starts with a king who once got angry with his beloved daughter and
cursed her that the no-good will take her. After she disappeared the king
became so sorry and regretful that his Second-to-The-King volunteered to
go out and look for the princess. After many hardships he found the
princess who instructed him how to save her by fasting, not sleeping and
longing for her, but twice he failed by eating or drinking and
consequently falling asleep and the princess was transferred to the golden
mountain. During his wanderings looking for the golden mountain he met a
giant who was responsible for all animals who tried to convince him to
give up, but even after all the animals did not know anything about the
golden mountain the Second-to-The-King remained determined to continue his
search. So the giant sent him to his brother who was responsible for all
the birds. After repeating the same scenario he sent the Second to the
third brother who was responsible for all the winds. After the winds did
not know about the golden mountain and the Second-to-The-King started to
cry, came another wind who apologized that it was late as it had to
deliver a princess to a golden mountain. The giant then ordered the wind
to take there also the Second and equipped him with a tool that will help
him there. The end of the story was not told; we only have a comment of
the editor saying that in the end he managed to save her but Rabbi Nachman
did not tell how.
        The story is usually interpreted along the lines described in the
first paragraph of this section. The king represents G-d and the pair
second-princess represents either the individual and his soul or the
Tzadik and the Divine Presence. The story was not completed by Rabbi
Nachman, because the story of creation is not finished and the harmonious
unification was not yet accomplished.

\section{Diagonal Links in "The Story of The Lost Princess"}
        When performing the diagonal analysis on "The Story of the Lost
Princess", 18 diagonal links were found. Two examples are presented in
table 2.
Our assumption was that diagonal links indicate key-events in the story
and in order to check if the events associated with these 18 diagonal
links were to be considered important events we went through 21 authors
who commentated "The Story of the Lost Princess". We looked at the 29
events they referred to and ranked them according to the number of writers
referring to them. When comparing the events pointed by the diagonal links
we found that they matched quite fully the top part of the list: the first
event that was not pointed by a diagonal link was number 14 in the ranked
list and only one of these events matched a low-ranked event (number 22).
The detailed analysis of this study will be published elsewhere. For the
purpose of this paper it is enough that appearance of a diagonal link
indicates a crucially significant point in the story.

\section{"The Lost Princess" and its parallel "The Raven" of the Grimm Brothers} 
        The Grimm Brothers' story "The Raven" starts with a queen who lost
her patience with her daughter and cursed her to become a raven, which she
did and flew to a black forest. A man going through the forest heard the
raven and went to it. He met the princess who instructed him how to save
her by fasting, not sleeping and waiting for her at a specific time.
Thrice he failed by drinking and falling asleep and the princess was
transferred to the golden palace. During his wanderings looking for the
golden palace he met a giant who first was hostile but then tried to help
him. He looked for the palace in his map but in vain. He tried a bigger
map and again - in vain. Then he convinced the man to wait for his brother
who, when came back, looked at his map that was still bigger but the
palace was not on it. At last they found the palace in an old map and the
giant took the man most of the way toward it. The man took from three
thieves the equipment that helped him enter the palace and release the
princess. The story ends with the princess thanking the man for saving her
and declaring their wedding to be celebrated the next day.
        This story is clearly a parallel to Rabbi Nachman's "The story of
the Lost Princess" although they are not identical. S. Pitrushka  claims
that neither of the two stories is the source of the other but they both
share a common source in the Polish story "A Diverting Story about the
Virtuous Princess Banialuce From an Eastern Land" by Hieronim Morsztyn
(died in 1655). Although Rabbi Nachman did not know the Polish language,
Pitrushka quotes the Polish historian A. Brinker who proved that "Princess
Banialuce" was translated to the Ukrainian language. 
        Finding a parallel story was indeed intriguing in light of the
fact that Rabbi Nachman did refer to the relations between his stories and
other folk tales. Rabbi Nachman's tales were edited by his main disciple
Rabbi Nathan Sternhartz of Nemirow. In his first introduction Rabbi Nathan
tells \cite{ref14}: "Before [the Rebbe (Hassidic sage)] began telling the
first story in this book, he declared, "Many hidden meanings and lofty
concepts are contained in the stories the world tell. These stories,
however, are deficient: they contain many omissions. They are also
confused, and people do not tell them in the correct order ...". The Rebbe
spoke about this. Then he began to tell the story on the following page.
He said, "I told this story while on a journey ..."
In contemporary words, Rabbi Nachman claims that folk tales need to be
upgraded and is moving to give an example how he does it. Therefore it was
especially interesting to perform a diagonal analysis on "The Raven" \cite{ref16}.

\section{Results}
In the analysis we found 12 diagonal links in "The Raven". Further
examination showed that out of the 18 diagonal links of "The Lost
Princess" only 9 were present in "The Raven". In 8 out of the 9 missing
appearances, the event itself was present in "The Raven" but its
description lacked the diagonal link. On the other hand, in all three
diagonal links that appeared in "The Raven" and did not appear in "The
Lost Princess" the event itself was not part of the story. Two of these
are placed in the last part of the story, the part that Rabbi Nachman did
not tell.
        When calculating the density of diagonal links' appearances
according to the number of words we found that in "The Raven" there were
12 diagonal links in 1652 words, which is a density of 0.7 percent. In "The Lost
Princess" there were 18 diagonal links in 1312 words, which is a density
of 1.4 percent.

\section{Meaning we attach to findings}
        In an earlier section we showed that an appearance of a diagonal link
indicates a part of the story which is of special quality. From this
obviously derives that the overall importance of the text increases with
the density of diagonal links appearances. The double density that was
found in "The Lost Princess" in comparison to "The Raven" indicates a
significant difference in quality. According to our findings Rabbi Nachman
managed to execute what he was trying to do - to upgrade the story.

\section{Future Experiments}
        Our intention now is to analyze the story of "Princess Banialuce"
and one of the theoretical teachings of Rabbi Nachman. According to the
above we expect to find a significant difference in the diagonal density
between "The Lost Princess" and "Princess Banialuce" and a non-significant
difference between the stories and the teachings of Rabbi Nachman.

Fig.1: A system which is a thin brick wall

Fig.2: Transitivity Connection

Fig.3: Component Connection Template as Diagonal Link

Tab.1: Example for the Component Connection Template

Tab.2: Two examples of diagonal links found in "The Story of the Lost
Princess"


\begin{thebibliography}  {10}
%
\bibitem{ref1}
{S. Solomon}, {\em The Microscopic Representation of Complex Macroscopic Phenomena}, Annual Reviews
of Computational Physics II pp 243-294 ({World Scientific}, 1995).
%
\bibitem{ref2}
{G.Mack}, {\em Pushing Einstein's Principles to the Extreme}, in {\em Quantum Fields and Quantum Space
Time}, edited by {G. 't
Hooft et al} ({Plenum Press}, 1997).
%
\bibitem{ref3}
{G.Mack}, {\em Universal Dynamics of Complex Adaptive Systems: Gauge Theory
of Things Alive}, DESY {\bf 94-075}, (April 1994).
%
\bibitem{ref4}
{G. Mack}, {\em Gauge Theory of Things Alive: Universal Dynamics as a Tool
in Parallel Computing}, Prog. Theor. Physics Suppl. {\bf 122}, 201 (1996).
%
\bibitem{ref5}
{N. Rubin, S. Hochstein and S. Solomon}, {\em {Restricted Ability to Recover 3D Global Motion from 1D
Motion Signals: Theoretical Observations}}, Vision
Research {\bf 35}, 569 (1995);
{N. Rubin, S. Hochstein and S. Solomon}, {\em { Restricted Ability to Recover 3D Global Motion from 1D
Motion Signals:
Psychophysical Observations}}, Vision
Research {\bf 35}, 463 (1995);
{S. Solomon}, {\em The Microscopic Bases of Vision}, in {Perspectives on
Cognitive Science}, Vol. 2, edited by {B. Kokinov} ({NBU}, 1995).
%
\bibitem{ref6}
{E. Adi, I. Levin and S.Solomon}, {\em The Dynamics of Childrens's Drawings}, in {Perspectives on
Cognitive Science}, Vol. 2, edited by {B. Kokinov} ({NBU}, 1995);
{E. Adi-Japha, I. Levin and S. Solomon}, {\em Emergence of Representation in Drawing: The Relation
Between Kinematic
and  Referential Aspect}, Cognitive Development {\bf 13}, 25 (1998).
%
\bibitem{ref7}
{N. Persky and S.Solomon}, {\em Collective Degrees of Freedom and Multiscale Dynamics in Spin Glasse},
Phys. Rev. E. {\bf 54}, 4399 (1996);
{I. Stenhill, S. Solomon and K. Wolowelsky}, {\em Dynamical Algebraic multi-grid simulations of free fields
on RTS}, Computer Physics Communications {\bf 83}, 152 (1994);
{E. Adi and S. Solomon}, {\em The solution to Wheeler-De Witt is Eight}, Phys.
Lett. B {\bf 336}, 152 (1994).
%
\bibitem{ref8}
{Y. Goldenberg, R. Horowitz, D. Mazursky and S. Solomon}, {\em New Product
Development With Prescribed Algorithms}, International Journal of Modern
Physics C, Vol. 8 No. 2, 365 (1996).
%
\bibitem{ref9}
{Y. Goldenberg, D. Mazursky and S. Solomon}, {\em Toward Identifying the
Inventive Templates of New Products: A Channeled Ideation Approach}, Journal of Marketing Research {\bf
36}, 200 (May 1999).
%
\bibitem{ref10}
{Y. Goldenberg, S. Solomon and D. Mazursky}, {\em Cluster Concepts Dynamics
Leading to Creative Ideas Without Critical Slowing Down}, International
Journal of Modern Physics C, Vol. 7 No. 5, 655 (1996).
%
\bibitem{ref11}
{Y. Goldenberg, D. Mazursky and S. Solomon}, {\em Creativity Templates:
towards Identifying the Fundamental Schemes of Quality Advertisements}, Marketing Science, Vol. 18, No. 3,
333 (1999).
%
\bibitem{ref12}
{Goldenberg Jacob and Mazursky David}, {\em {Creativity Templates in Product
innovation}} ({Cambridge University Press}, (In Press)).
%
\bibitem{ref13}
{Band, Arnold J.}, {\em {Nachman of Bratslav - The Tales}} ({Paulist Press}, {New York} 1978).
%
\bibitem{ref14}
{Kaplan, Rabbi Aryeh}, {\em {Rabbi Nachman's Stories}} ({Breslov Research
Institute}, {New York} 1983).
%
\bibitem{ref15}
{S. Pitruska}, {\em {Polish Source for Rabbi Nachman of Brastlav's Tales}}, Ktuvim, Year B, Issue 41-42,
12.7.1928 (In Hebrew).
%
\bibitem{ref16}
{In order to allow further examination and refinement we attach the
URLs of the full texts of two stories:
http://www.breslov.org/princess.html  and:
http://www.cs.cmu.edu/~spok/grimmtmp/070.txt}
%
\end{thebibliography}
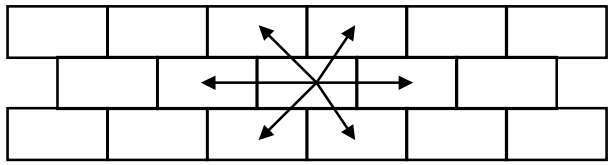
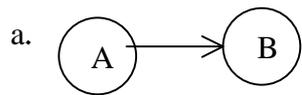
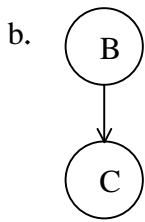
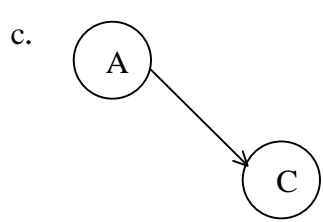
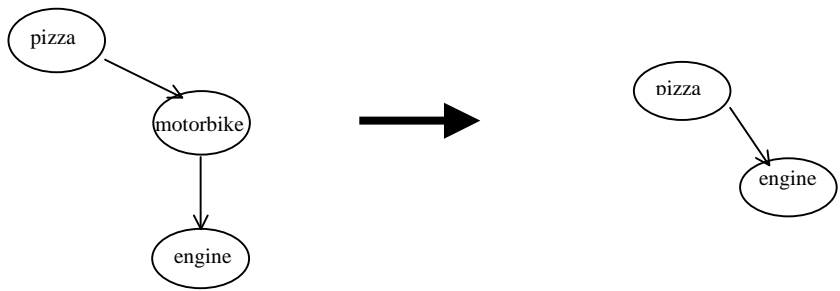
\end{document}